\documentclass[twocolumn,prb,aps,amsmath,amssymb]{revtex4}

\usepackage{graphicx}% Include figure files
\usepackage{dcolumn}% Align table columns on decimal point
\usepackage{bm}% bold math

\begin{document}

%\preprint{APS/123-QED}

\title{Enhanced critical current density of YBa$_2$Cu$_3$O$_x$ films grown on 
Nd$_{1/3}$Eu$_{1/3}$Gd$_{1/3}$Ba$_2$Cu$_3$O$_x$ with nano-undulated surface morphology}

\author{R.~L.~Meng$^{1}$}
\author{T.~H.~Johansen$^{1,2}$}
\author{I.~A.~Rusakova$^{1}$}
\author{A.~Baikalov$^{1}$}
\author{D.~Pham$^{1}$}
\author{F.~Chen$^{1}$}
\author{Z.~Y.~Zuo$^{1}$}
\author{C.~W.~Chu$^{1,3,4}$}
\affiliation{$^{1}$Department of Physics and Texas Center for Superconductivity, 
University 
of Houston, 202 Houston Science Center, Houston, Texas 77204-5002}
\affiliation{$^{2}$Department of Physics, University of Oslo, N-0316 Oslo, Norway}
\affiliation{$^{3}$Lawrence Berkeley National Laboratory, 1 Cyclotron Road, 
Berkeley, California 94720}
\affiliation{$^{4}$Hong Kong University of Science and Technology, 
Hong Kong}

\date{\today}% It is always \today, today,

\begin{abstract}
We report a simple and easily controllable method where a nano-undulated surface morphology of 
Nd$_{1/3}$Eu$_{1/3}$Gd$_{1/3}$Ba$_2$Cu$_3$O$_x$ (NEG) films leads to a substantial increase in the 
critical current density in superconducting YBa$_2$Cu$_3$O$_x$ (YBCO) films deposited by pulsed 
laser deposition on such NEG layers. The enhancement is observed over a wide range of fields and 
temperatures. Transmission electron microscopy shows that such YBCO films possess a high density 
of localized areas, typically $20 \times 20$ nm$^2$ in size, where distortion of atomic planes give 
rotational (2 to 5$^\circ$) moir\'e patterns. Their distribution is random and uniform, and expected 
to be the origin of the enhanced flux pinning. Magneto-optical imaging shows that these films have 
excellent macroscopic magnetic uniformity.
\end{abstract}

\maketitle

Practical applications of high temperature superconductor films depend crucially upon finding ways 
to enhance the flux pinning and thereby increasing the critical current density, $j_c$, especially at 
high magnetic fields. Recent reports have shown that pre-decoration of the substrate by a high 
density of nano-sized particles is an efficient way of creating large numbers of strong pinning 
sites in the superconducting film that subsequently is deposited on the decorated surface. The basic 
idea of the method is using the nano particles to create a substantial lattice mismatch or chemical 
poisoning so that locally the superconducting phase is prevented to form. Successful examples of 
this are sputtering nano-dots of Ag on a SrTiO$_3$ (STO) substrate prior to deposition of 
(Cu,Tl)BaSrCa$_2$Cu$_3$O$_y$, pulsed laser deposition of nano-islands of Y$_2$O$_3$ or Ag on STO and 
YSZ substrates, respectively, prior to deposition of YBa$_2$Cu$_3$O$_x$ (YBCO).\cite{cri,mat,ion} 
In principle, the method can be extended by repeating the double deposition, as was demonstrated 
with alternating growth of an ultra thin layer of second-phase YBa$_2$CuO$_5$ and superconducting 
YBCO repeated up to 200 times.\cite{hau}

In this work we report a new and efficient method to obtain enhanced pinning in films of YBCO. 
The method is based on our observation that thin films of the mixed rare earth compound 
Nd$_{1/3}$Eu$_{1/3}$Gd$_{1/3}$Ba$_2$Cu$_3$O$_x$ (NEG) grown by laser ablation on STO substrates 
develop a surface morphology with densely packed and sharply separated submicron sized growth 
islands. We show that by using such a nano-undulated surface as a sublayer for deposition of 
YBCO films one obtains an increase in $j_c$ by approximately 50\%. Magneto-optical (MO) imaging 
studies reveal that such YBCO films have excellent uniformity, and are therefore well suited 
for device applications. Moreover, since the NEG sublayer itself is superconducting, the method 
also gives a high engineering $j_c$. 

Targets of NEG were prepared with stoichiometric R$_2$O$_3$ (R = Nd, Eu, Gd), BaCO$_3$ and CuO 
powders sintered at 950 $^{\circ}$C. X-ray diffraction confirmed that the target consists of pure 
123 phases. The NEG films were deposited on single crystal (001) STO substrates by pulsed laser 
deposition. Before deposition the substrates were cleaned by heating to 900 $^{\circ}$C for 
30 minutes. The films were deposited at a temperature of 810--830 $^{\circ}$C in a 350 mTorr 
oxygen atmosphere using a KrF excimer laser with RF power of 250 mJ. 

\begin{figure}
\includegraphics{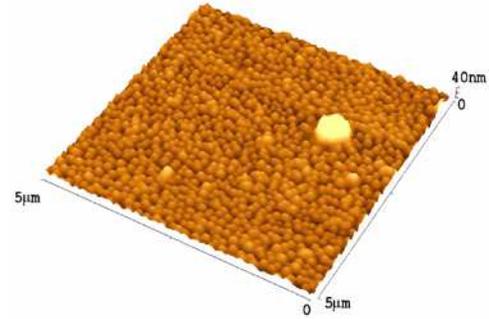}
\caption{\label{fig:fig1}AFM image of a $5 \times 5$ $\mu$m$^2$ surface area of a NEG film deposited 
on STO, serving as sub-layer for subsequent deposition of a YBCO film. The NEG growth granules are 
nearly mono-disperse with diameter 80--100 nm and height 20--25 nm.}
\end{figure}

Shown in Fig.~\ref{fig:fig1} is the surface morphology of a typical bare NEG film observed using 
atomic force microscopy (AFM). This 100 nm thick film is densely packed with growth islands 
resulting in an undulated surface having a highly uniform and narrow distribution of peaks 
15--25 nm high and 80--100 nm in diameter. This type of surface morphology is similar to that 
reported by Cai et al.\cite{cai}, and also quantitatively the AFM results are in good agreement. 

In synthesizing the two-layer films the deposition of YBCO and NEG was done in the same process. 
We found that optimal conditions for YBCO deposition is to use the same oxygen pressure and laser 
energy as for NEG, and lowering the deposition temperature close to 800 $^{\circ}$C. After 
deposition, the films were in-situ annealed at 450--500 $^{\circ}$C maintaining the oxygen 
pressure for 30 minutes, before cooling down to room temperature. No ex-situ annealing was employed. 
Note that the synthesis of the two films in the proper order is possible because the melting point 
of NEG is the higher of the two compounds.

For comparison, films of YBCO were also deposited directly on STO substrates using the same 
conditions. Transport measurements showed a transition temperature of 92 K for the YBCO films. 
The film thickness was measured using $\alpha$-step surface profilometry.

\begin{figure}
\includegraphics{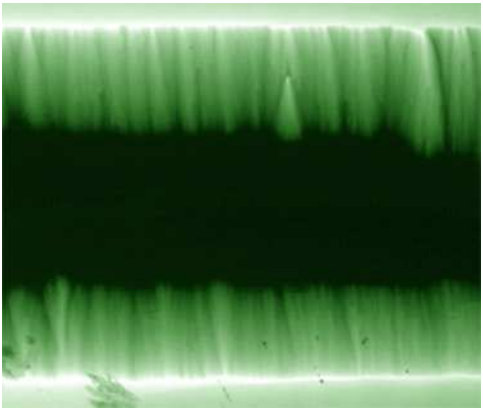}
\caption{\label{fig:fig2}MO image of flux penetration in a two-layer YBCO/NEG film. The image was 
recorded at 5 K in a perpendicular applied field of 45 mT, and shows the strip shaped sample in a 
partially penetrated state. The dark central band is the Meissner state part of the film, and the 
field exclusion causes the enhanced brightness seen along the film edge. The strip width is 3 mm.}
\end{figure}

All samples were investigated by MO imaging using in-plane magnetized bismuth substituted iron 
garnet films as indicator. The setup consists of an Olympus polarizing microscope and an Oxford 
Microstat-He optical cryostat with a custom made coil to apply an external magnetic field. We used 
a fully crossed polarizer and analyzer setting, giving images where the brightness represents the 
magnitude of the local flux density. Shown in Fig.~\ref{fig:fig2} is an MO image of a YBCO/NEG 
sample with layer thicknesses of 100 nm and 50 nm, respectively. The image was taken at 5 K in 
an applied field of $B_a = 45$ mT. As seen directly from the image, the two-layer film has 
excellent uniformity in superconducting properties on the macroscopic scale. Only one defect in 
the upper half of the film is visible, as it creates a parabolic fan-like flux pattern starting 
from a point inside the strip. Since the superconducting film covers the whole substrate area, 
a slight edge roughness is also causing fan-like flux structures, which can be seen starting 
from both the upper and lower edge in the image.

From MO images the low-field critical current density was determined from the Bean model formula 
for a long thin strip, $\mu_{0}J_{c}=\pi B_{a}/\cosh^{-1}(w/a)$. Here $J_c$ is the sheet current 
(the current density integrated over the film thickness), and $a$ and $w$ are the width of the 
central flux free area and the width of the strip itself, respectively. Since these samples consist 
of two different superconducting layers, the sheet current has two contributions; 
$J_{c} = j_{c}d + j_{c}^{NEG}d^{NEG}$, where $j_c$ and $d$ are the critical current density and 
thickness of the YBCO layer, and where the second term represents the current flowing in the NEG 
layer. We determined $j_{c}^{NEG}$ from MO images of flux penetration in bare NEG films prepared 
under the same deposition conditions. The current density in this layer was not very high, 
e.g., $j_{c}^{NEG}$(5 K) $= 0.7 \cdot 10^7$ A/cm$^2$, but then optimizing the critical current in 
the NEG film was not our focus in the present work. Using the procedure described above we find 
for the YBCO layer alone that $j_{c} = 7.1 \cdot 10^7$ A/cm$^2$ at 5 K. At higher temperatures we 
obtain the values listed in the Table~\ref{tab:table1}. Included in the table is also $j_{c}$ values measured on 
a reference YBCO/STO sample grown under the same conditions. We find consistently that YBCO on 
NEG gives an enhancement in $j_c$ of 50\%--100\% between 5 K and 77 K.

\begin{table}
\caption{\label{tab:table1}The table shows $j_c$ in units of $10^7$ A/cm$^2$ for the YBCO film
grown on a NEG sublayer at different temperatures. Shown is also the $j_c$ of a reference 
YBCO film.}
\begin{ruledtabular}
\begin{tabular}{cccccccc}
 &5 K&60 K&77 K\\
\hline
YBCO/NEG/STO& 7.2 & 1.8 & 0.46 \\
YBCO/STO& 4.7 & 1.2 & 0.24 \\
\end{tabular}
\end{ruledtabular}
\end{table}

\begin{figure}
\includegraphics{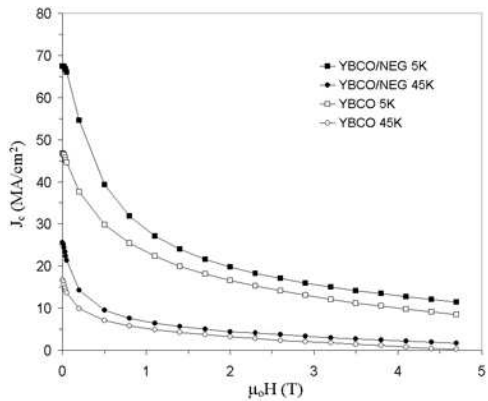}
\caption{\label{fig:fig3}Critical current density in YBCO films as a function of applied magnetic 
field. The solid symbols represent YBCO deposited on a NEG sub-layer, and open symbols show YBCO 
on the bare STO substrate.}
\end{figure}

The field dependence of $j_{c}$ was measured in the field range from zero and up to 5 T using a 
SQUID magnetometer. The results are plotted in Fig.~\ref{fig:fig3}, where full symbols show 
$j_{c}$ of YBCO on NEG, and open symbols represent YBCO/STO. For the two-layer film $j_{c}$ of 
the YBCO part was extracted using that in fully penetrated states the measured magnetic moment 
equals $m = (j_{c}d + j_{c}^{NEG}d^{NEG})$ $\times$ geometrical factor, where the second factor is 
given by the sides of the  rectangular sample used for the SQUID measurements. The results clearly 
show that also the field behavior of the YBCO film is largely improved by the NEG sublayer. Over 
the whole field range $j_{c}$ is increased by 40--50\% both at 5 K and 45 K. Note that the zero-field 
$j_{c}$ obtained from the M-H loop width is slightly lower than the values obtained from MOI, 
which is to be expected as explained in Ref. \onlinecite{sha}. 

\begin{figure}
\includegraphics{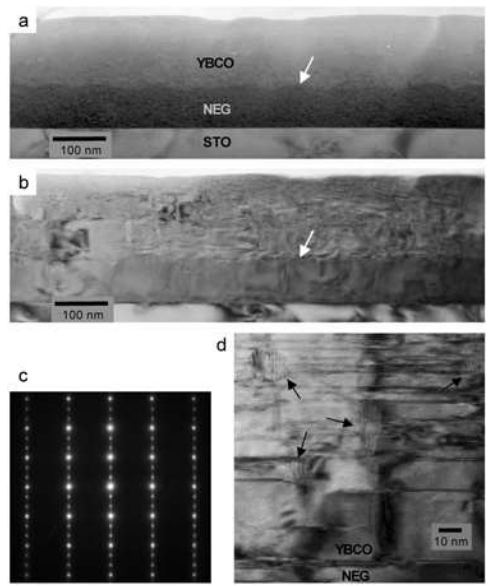}
\caption{\label{fig:fig4}(a) TEM image of the two-layer film cross section taken under mass contrast 
imaging conditions. (b) Conventional TEM bright-field image of the interface. In both panels, 
the arrow points at the undulated YBCO/NEG interface. (c) Selected area diffractogram representing 
the full YBCO/NEG/STO structure. (d) High resolution TEM of the YBCO layer showing several areas 
(see arrows) with rotational moir\'e patterns.}
\end{figure}

To clarify the origin of this pinning enhancement the samples were investigated by transmission 
electron microscopy (TEM) using a JEOL 2000FX microscope operated at 200 kV. Cross-sectional 
samples were prepared by a standard procedure and ion milling was carried out with 4 keV Ar ions. 
Shown in Fig.~\ref{fig:fig4}a is a TEM bright-field image of a double layer film obtained under 
mass-thickness contrast image formation conditions. This type of contrast arises from incoherent 
(Rutherford) elastic scattering of electrons, which is a strong function of atomic number $Z$. The 
difference between the Y ions ($Z = 39$) in the upper YBCO film and the much heavier ions of 
Nd, Eu, and  Gd ($Z = 60$, 63, and 64, respectively) in the sublayer results in a clear contrast 
between the two films. Their interface has a very distinct wavy appearance, which is in full 
quantitative agreement both in inter-peak distance and in undulation amplitude with the AFM image 
obtained for the bare NEG film. We conclude therefore that its surface morphology remains intact 
throughout the deposition of the YBCO film. Note also from the TEM image that both the NEG/STO and 
the YBCO/NEG interfaces are very uniform. Moreover, selected area electron diffraction (SAED) 
recorded from the substrate and the two layers shows that both films are very well c-axis aligned 
with the substrate, see Fig.~\ref{fig:fig4}c.

Shown in Fig.~\ref{fig:fig4}b is the microstructure of the double layer film obtained by 
conventional bright-field TEM revealing strain contrast. While the NEG/STO shows strain mainly 
along the interface, the YBCO layer contains numerous strained regions throughout its volume. 
Whereas the strain contrast at the NEG/STO interface is due to mismatch of lattice parameters, 
this is not the case for the YBCO/NEG interface which has nearly identical lattice constants. 
This strongly suggests that the strained regions inside the YBCO film stem from the interface 
undulation. Indeed, high resolution TEM, see Fig.~\ref{fig:fig4}d, reveals that in the YBCO layer 
strained regions start from the interface. Moreover, we find that locally the atomic planes have 
orientation deviations from 2 to 5$^{\circ}$ resulting in rotational moir\'e patterns, four of them 
shown by arrows. The size of the moir\'e pattern areas is very small, typically 
$20 \times 20$ nm$^2$, and they are randomly and quite uniformly distributed. We believe that the 
presence of these structural defects strongly contributes to the enhancement of the flux pinning, 
and thereby the critical current density in such YBCO films. Additional contributions can come 
from the high density of stacking faults, and therefore partial dislocations, as also found to be 
present in these films.

In conclusion, we have demonstrated that a nano-undulated surface morphology of NEG films leads to 
a substantial increase in the critical current density in YBCO films deposited on top of the NEG 
layer. The enhancement is observed over a wide range of fields and temperatures. Compared to most 
other methods of nano-patterning of substrates, this new method is technologically very simple, 
easily controllable and fast, and also economically favorable. An interesting extension of this 
work would be to make a periodic multilayer structure YBCO/NEG\dots/YBCO/NEG/STO to provide films with 
large total critical current. Since the NEG sublayer is also superconducting the engineering 
critical current of such a structure could then become very high.

\begin{acknowledgments}
We thank Y.~Y.~Sun for X-ray analysis of the samples. The work in Houston is supported in part by 
NSF Grant No. DMR-9804325, the T.~L.~L. Temple Foundation, the John J. and Rebecca 
Moores Endowment, and the State of Texas through the Texas Center for 
Superconductivity at the University of Houston; and at Lawrence Berkeley 
Laboratory by the Director, Office of Science, Office of Basic Energy Sciences, 
Division of Materials Sciences and Engineering of the U.S. Department of Energy 
under Contract No. DE-AC03-76SF00098. 
One of the authors (THJ) is grateful to the Norwegian Research Council for financial support.
\end{acknowledgments}

%\bibliographystyle{plain}

\begin{thebibliography}{99}

\bibitem{cri} A. Crisan, S. Fujiwara, J.C. Nie, A. Sundaresan and H. Ihara, Appl. Phys. Lett. 
{\bf 79}, 4547 (2001).
\bibitem{mat} K. Matsumoto et al., Physica C {\bf 412--414}, 1267 (2004).
\bibitem{ion} M. Ionescu, A.H. Li, Y. Zhao, H.K. Liu and A. Crisan, J. Phys. D: Appl. Phys. {\bf 37}, 
1824 (2004)
\bibitem{hau} T. Haugan, P.N. Barnes, R. Wheeler, F. Meisenkothen and M. Sumption, Nature {\bf 430}, 
867 (2004).
\bibitem{cai} C. Cai, B. Holzapfel, J. HŠnisch, L. Fernandez and L. Schultz, Phys. Rev. B {\bf 69}, 
104531 (2004).
\bibitem{sha} D. V. Shantsev, Y. M. Galperin and T. H. Johansen, Phys. Rev. B {\bf 61}, 9699 (2000).
\end{thebibliography}

\end{document}